\documentclass[11pt]{article}
\usepackage{amsfonts,amsmath}

\newtheorem{lemma}{Lemma}
\newtheorem{theorem}{Theorem}

\def\res{{\mathrm{res}}}
\def\beq{\begin{equation}}
\def\eeq{\end{equation}}
\def\R{{\mathbb R}}

\def\C{{\mathbb C}}

\begin{document}

\title{On some algebraic examples of Frobenius manifolds
\thanks{The work was
supported by RFBR (no. 06-01-00094a) and the complex integration
project 2.15 of SB RAS. The first author (A.E.M.) was also
supported by the grant of President of Russian Federation (grant
MK-9651.2006.1) and Russian Science Support Foundation. }}
\author{A.E. Mironov \thanks{Institute of Mathematics,
630090 Novosibirsk, Russia; mironov@math.nsc.ru} \and I.A.
Taimanov \thanks{Institute of Mathematics, 630090 Novosibirsk,
Russia; taimanov@math.nsc.ru}}
\date{}
\maketitle

\section{Introduction}

In this paper we demonstrate how to construct explicit examples of
Frobenius manifolds by using analytical methods of finite-gap
integration. Therewith we apply Krichever's scheme of constructing
solutions to the associativity equations \cite{K}. Although it is
rather clear that the solutions to associativity equations
corresponding to smooth spectral curves are not quasihomogeneous
we show that in a very degenerate case when the spectral curve
consists of rational irreducible components one may construct
quasihomogeneous solutions to these equations. The extension of
these solutions to Frobenius manifolds is achieved by using some
technical algebraic lemma which is exposed in \S
\ref{algebraiclemma}.

Until recently all known Frobenius manifolds were given by
original Dubrovin's examples of Frobenius structures on the spaces
of orbits of the Coxeter groups (in this case Dubrovin used the
Saito flat metric on the space of orbits and such solutions to the
WDVV equations corresponding to the $A_n$ singularities were found
in \cite{DVV}) and on the Hurwitz spaces, by quantum cohomology,
and by the extended moduli space of complex structures on
Calabi--Yau manifolds \cite{BK}. In \cite{Sh} this list was
expanded by Shramchenko who ``doubled'' Frobenius structures by
Dubrovin on the Hurwitz spaces (Shramchenko's manifolds have twice
the dimension of the Hurwitz spaces).

In all these cases the manifold with such a structure has its own
specified geometrical meaning and only quantum cohomology can be
not semisimple, i.e. contain nilpotent elements in a tangent
Frobenius algebra at a generic point. Our examples always lack the
semisimplicity property (thus they are not directly related to
isomonodromic deformations, see \cite{D}) and are obtained by
analytical methods without any recognition of their relations to
other geometrical objects. These examples are algebraic in the sense that the
correlators $c_{ijk} = \frac{\partial^3 F}{\partial x^i \partial x^j \partial x^k}$
are algebraic functions.

\section{Some preliminary facts on Egoroff metrics and Frobenius
manifolds}

Given a symmetric tensor $\eta^{\alpha\beta} =
\eta^{\beta\alpha}$, the associativity equations for the function
$F$ take the form \beq \label{e1} \frac{\partial^3 F(t)}{\partial
t^\alpha \partial t^\beta \partial t^\lambda} \, \eta^{\lambda
\mu} \frac{\partial^3 F(t)}{\partial t^\gamma \partial t^\delta
\partial t^\mu} = \frac{\partial^3 F(t)}{\partial t^\gamma
\partial t^\beta \partial t^\lambda} \, \eta^{\lambda \mu}
\frac{\partial^3 F(t)}{\partial t^\alpha \partial t^\delta
\partial t^\mu}, \eeq where $t=(t^1,\dots,t^n)$ and the indices
range from $1$ to $n$. They are equivalent to the condition that
the finite-dimensional algebra with generators $e_1,\dots,e_n$ and
the commutative multiplication
$$
e_\alpha \cdot e_\beta = c^\gamma_{\alpha \beta} e_\gamma, \ \ \
c_{\alpha \beta \gamma} = \frac{\partial^3 F(t)}{\partial t^\alpha
\partial t^\beta \partial t^\gamma}, \ \ \ c^\gamma_{\alpha
\beta}=\eta^{\gamma\delta}c_{\alpha \beta \delta},
$$
is associative with respect to the multiplication, i.e. we have
$$
(e_\alpha \cdot e_\beta)\cdot e_\gamma = e_\alpha \cdot (e_\beta \cdot e_\gamma)
\ \ \ \mbox{for all $\alpha,\beta,\gamma$.}
$$

These equations first appeared in the topological field theory where
together with conditions
$$
c_{1\alpha\beta} = \eta_{\alpha\beta}, \ \ \alpha,\beta=1,\dots,n; \ \ \
\eta^{\alpha\beta}\eta_{\beta\gamma} = \delta^\alpha_\gamma,
$$
with $\eta_{\alpha\beta}$ a constant metric, probably indefinite,
and
\beq
\label{e2}
F(\lambda^{d_1}t^1,\dots,\lambda^{d_n}t^n) =
\lambda^{d_F}F(t^1,\dots,t^n)
\eeq
(the quasihomogeneity
condition) they the system of
Witten--Dijkgraaf--Ver\-linde--Verlinde (WDVV) equations
\cite{W,DVV}.

The quasihomogeneity condition is generalized as
follows: it is assumed that there is the vector field $E =
(q^\alpha _\beta t^\beta + r^\alpha) \partial_\alpha$ such that
$E^\alpha \partial_\alpha F = d_F F$ (in the case of (\ref{e2}) we
have $E = d_1 t^1 \partial_1 + \dots + d_n t^n
\partial_n$) and this generalization covers the case of quantum cohomology.

Since, by \cite{DVV}, it is only important for the correlators
$c_{ijk}$, i.e. third derivatives of $F$, to be quasihomogeneous in the sense of (\ref{e2}) there
is another generalization of quasihomogeneity which reads that
$$
E^\alpha \partial_\alpha F = d_F F + (\mbox{a polynomial of second order in}\ t^1,\dots,t^n).
$$
This generalization is important for us because in our examples part of exponents
$d_i$ equal to $-1$.

The geometric counterpart of a solution to the WDVV equations is a
Frobenius manifold which notion was introduced by Dubrovin
\cite{D} who discovered rich differential-geometrical properties
of the WDVV equation and thus gave rise to the Frobenius geometry.

There is an important relation between Frobenius manifolds and Egoroff metrics also
discovered by Dubrovin \cite{D0}.

A metric
\beq
\label{e3}
ds^2 = \sum_{i=1}^n H^2_i(u) \left(du^i\right)^2
\eeq
is called Egoroff if the rotation coefficients $\beta_{ij} = \frac{\partial_i H_j}{H_i}, i\neq j$,
are symmetric: $\beta_{ij}=\beta_{ji}$.
Let us consider the Darboux--Egoroff metrics, i.e., flat Egoroff metrics
$$
\eta_{\alpha\beta} dx^\alpha dx^\beta = \sum_{i=1}^n H^2_i(u) \left(du^i\right)^2
$$
where $x^1,\dots,x^n$ are flat coordinates in some domain where the
coefficients $\eta_{\alpha\beta}$ are constant. We have
$\eta^{\alpha\beta} = \sum_i H^{-2}_i \frac{\partial
x^\alpha}{\partial u^i} \frac{\partial x^\beta}{\partial u^i}$ and
the flatness condition together with symmetry of the rotation
coefficients imply that there is a function $F$ called the
prepotential such that
\beq
\label{e4}
c_{\alpha \beta \gamma} =
\sum_i^n H_i^2 \frac{\partial u^i}{\partial x^\alpha} \frac{\partial
u^i}{\partial x^\beta}\frac{\partial u^i}{\partial x^\gamma} =
\frac{\partial^3 F}{\partial x^\alpha \partial x^\beta \partial
x^\gamma}
\eeq
and the associativity equations hold:
$$
c^\lambda_{\alpha \beta} c^\mu_{\lambda\gamma} =
c^\mu_{\alpha\lambda} c^\lambda_{\beta\gamma} \ \ \ \mbox{for all} \
\alpha,\beta,\gamma=1,\dots,n,
$$
where
$$
c^\alpha_{\beta\gamma} = \sum_i \frac{\partial x^\alpha}{\partial u^i}
\frac{\partial u^i}{\partial x^\beta}\frac{\partial u^i}{\partial x^\gamma}.
$$
The inverse is also true assuming that this associative algebra is
semisimple: one may construct from such a solution $F(t)$ to the
associativity equations a Egoroff metric meeting (\ref{e4}).

\section{Finite gap construction of Egoroff metrics and Frobenius manifolds}

The condition that the formula (\ref{e3}) defines the Euclidean metric
$ds^2 = $ $ = \eta_{\alpha\beta}dx^\alpha dx^\beta = \delta_{\alpha\beta}dx^\alpha dx^\beta$
in some domain (without assuming that the rotation coefficients are symmetric)
means that $u^1,\dots,u^n$ are curvilinear $n$-orthogonal coordinates in this domain and
it is written in the form of the Darboux equations.

First the methods of integrable systems were applied for constructing explicit solution to
the Darboux system by Zakharov \cite{Z} who used the dressing method and then
this approach was extended by Krichever onto the finite gap integration method \cite{K}.

In \cite{MT} we already applied Krichever's procedure to a very
degenerate case when the spectral curve is reducible and all its
reducible components are rational. In this case the procedure of
constructing solutions reduces to linear equations.

We consider the same spectral curves in this paper.

Let $\Gamma$ be a reducible algebraic curve such that every of its
irreducible components $\Gamma_1,\dots,\Gamma_s$ is isomorphic to
${\mathbb C}P^1$ and all singularities on $\Gamma$ are
intersections of different components.

A regular differential $\Omega$ on $\Gamma$ is defined by
meromorphic differentials $\Omega_1,\dots,\Omega_s$ on the
components such that every such a differential may have poles only
simple poles and only at the intersection points of the components
and the sum of the residues at every intersection point vanishes:
$\sum_{j=1}^r {\rm res_P}\Omega_{i_j}=0, \ \ \ P \in \Gamma_{i_1}
\cap \dots \cap \Gamma_{i_r}$.

Let us take three divisors on $\Gamma$:
$$
P = P_1 + \dots + P_n, \ \
D = \gamma_1 + \dots + \gamma_{g_a+l-1}, \ \
R = R_1 + \dots + R_l,
$$
where $g_a$ is the arithmetic genus of $\Gamma$.
Let us denote by $k_i^{-1}$ some local parameter near $P_i$, $i=1,\dots,n$.
It is said that $\psi(u^1,\dots,u^n,z),\ z\in\Gamma$, is the
Baker--Akhiezer function corresponding to the data
$S=\{P,D,R\}$
if

1) $\psi\exp(- u^i k_i)$ is analytic near $P_i$, $i=1,\dots,n$;

2) $\psi$ is meromorphic on $\Gamma\backslash\{\cup P_i\}$ with
poles at $\gamma_j$, $j=1,\dots,g_a+l-1$;

3) $\psi(u,R_k)=1$, $k =1,\dots,l$.

Let us take an additional divisor
$Q=Q_1+\dots+Q_n$ on $\Gamma$ such that $Q_i \in \Gamma \setminus
\{P \cup D \cup R\}, i=1,\dots,n$ and put
$$
x^j(u^1,\dots,u^n)=\psi(u^1,\dots,u^n,Q_j), \ j =1,\dots,n.
$$

For such curves the Krichever scheme works as follows \cite{MT}:

\begin{itemize}
\item
Let $\Gamma$ admit a holomorphic involution
$\sigma:\Gamma\rightarrow\Gamma$ such that

1) $\sigma$ has exactly $2m, m \leq n$, fixed points
which are just $P_1,\dots, P_n \in P$ and $2m-n$ points from $Q$;

2) $\sigma(Q)=Q$, i.e. the non-fixed points from $Q$ are interchanged by the involution:
$$
\sigma(Q_k) = Q_{\sigma(k)}, \ \ \ k=1,\dots,n;
$$

3) $\sigma(k_i^{-1}) = -k_i^{-1}$ near $P_i$,
$i=1,\dots,n$;

4) there exists a regular differential $\Omega$ on $\Gamma$ such that
its divisors of zeros and poles have the form
$$
(\Omega)_0= D + \sigma D +P, \ \ \ (\Omega)_{\infty}=R+\sigma R+Q.
$$

Then $\Omega$ is a pullback of some meromorphic differential
$\Omega_0$ on $\Gamma_0 = \Gamma/\sigma$ and we have
$$
\sum_{k,l} \eta_{kl}\partial_{u^i}x^k\partial_{u^j} x^l =
\varepsilon^2_i h^2_i \delta_{ij},
$$
where
$$
h_i = \lim_{P \to P_i} \left(\psi e^{-u^i k_i}\right), \ \
\ \eta_{kl} = \delta_{k,\sigma(l)} \res_{Q_k} \Omega_0,
$$
and
$$
\Omega_0 = \frac{1}{2} \left(\varepsilon_i^2 \lambda_i +
O(\lambda_i)\right) d\lambda_i, \ \lambda_i = k_i^{-2}, \ \mbox{at
$P_i$, $i=1,\dots,n$}.
$$

Moreover if there is an antiholomorphic involution $\tau: \Gamma
\to \Gamma$ such that all fixed points of $\sigma$ are fixed by
$\tau$ and
$$
\tau^\ast(\Omega) = \overline{\Omega}
$$
then the coefficients $H_i(u)$ are real valued for $u^1,\dots,u^n
\in \R$  and $u^1,\dots,u^n$ are $n$-orthogonal coordinates in the
flat $n$-space with the metric $\eta_{kl} dx^k dx^l$.
\end{itemize}

The proof of this statement is basically the same as Krichever's original proof for the case of smooth spectral curves
\cite{K}. It is only necessary to consider regular differentials instead of meromorphic and specialize for $g$
the arithmetic genus which is different from the geometric genus for singular curves.

The following theorem distinguishes some special case when this
construction leads to Darboux--Egoroff metrics and
quasihomogeneous solutions to the associativity equations.

\begin{theorem}
1) Let every component $\Gamma_i, i=1,\dots,n$,
contain a pair of points $P_i = \infty,\ Q_i=0$ and $k_i^{-1} = z_i$ be a global parameter
on $\Gamma_i$. Let us also assume that any intersection point $a \in \Gamma_i \cap \Gamma_j$ of different components
has the same coordinates on both components:
$$
z_i(a)=z_j(a)
$$
and the involution $\sigma$
takes the form
$$
 \sigma(z_i)=-z_i.
$$
Then the metric
$$
ds^2 = \eta_{kl} d x^k d x^l = \sum_{i} \left(\varepsilon^2_i h^2_i\right) \left(du^i\right)^2, \ \
h_i=h_i(u^1,\dots,u^n), \ i=1,\dots,n,
$$
constructed from these spectral data is a Darboux--Egoroff metric.

2) Moreover assume that the spectral curve is connected and the
Ba\-ker--Akhiezer function is normalized just at one point $r$:
$$
\psi(u,r)=1, \ \ \ R = r \in \Gamma.
$$
Then the functions
$$
c_{\alpha \beta \gamma}(x)=\sum_{i=1}^n H_i^2\frac{\partial
u^i}{\partial x^\alpha} \frac{\partial u^i}{\partial
x^\beta}\frac{\partial u^i}{\partial x^\gamma}, \ \ \ H_i =
\varepsilon_i h_i,
$$
are homogeneous
$$
c_{\alpha\beta\gamma}(\lambda x^1,\dots,\lambda x^n)=
\frac{1}{\lambda}c_{\alpha\beta\gamma}(x^1,\dots,x^n).
$$
\end{theorem}

{\it Proof} of the first statement follows Krichever's scheme
\cite{K}. We take the meromorphic function $f:\Gamma \to \C$
defined by the parameters $z_i,i=1,\dots,n$, on the components:
$$
f(w) = z_i(w) \ \ \ \mbox{for $w \in \Gamma_i$}.
$$
Then the differential
$$
\omega = f(z) \frac{\partial_i \psi(u,z)}{h_i(u)} \frac{\partial_j \psi(u,\sigma(z))}{h_j(u)}
$$
has poles only at $P_i$ and $P_j$ with the residues $\beta_{ij}$ and $-\beta_{ji}$ which implies
$$
\sum {\rm res}\, \omega = \beta_{ij}-\beta_{ji}=0.
$$

Proof of the second statement immediately follows from Lemmata 1
and 2.

\begin{lemma}
Under the assumptions of Theorem 1, we have the equality
$$
x^j(u^1+\mu,\dots,u^n+\mu) = e^{-r\mu}x^j(u^1,\dots,u^n).
$$
\end{lemma}

{\it Proof.} On the component $\Gamma_j$ the function equals
$$
\psi_j (z_j) =
e^{u^jz_j}\left(f_{j0}(u)+\frac{f_{j1}(u)}{z_j-\gamma_1^j}+\dots
 +\frac{f_{jk_j}(u)}{z_j-\gamma_{k_j}^j}\right).
$$
Let $r \in \Gamma_p$. Then the condition $\psi(r)=1$ is written as
\beq \label{e5} f_{p0}(u)+\frac{f_{p1}(u)}{r-\gamma_1^p}+\dots
 +\frac{f_{pk_p}(u)}{r-\gamma_{k_p}^p}=e^{-ru^p}.
\eeq
If the components $\Gamma_i$ and $\Gamma_j$ intersect at some
point $a$ then this points has the same coordinates on both
components and the condition
$$
 \psi_j(a)=\psi_i(a),
$$
takes the form
\beq
\label{e6}
\begin{split}
 e^{a(u^j-u^i)}\left(f_{j0}(u)+\frac{f_{j1}(u)}{a-\gamma_1^j}+\dots
 +\frac{f_{jk_j}(u)}{a-\gamma_{k_j}^j}\right)=
\\
 \left(f_{i0}(u)+\frac{f_{i1}(u)}{a-\alpha_1^i}+\dots
 +\frac{f_{ik_i}(u)}{a-\alpha_{k_i}^i}\right).
\end{split}
\eeq
By (\ref{e5}) and (\ref{e6}), the translation
$$
u^j\rightarrow u^j+\mu,
$$
results in the multiplication of the coefficients $f_{sk}$:
$$
 f_{sk}\rightarrow f_{sk}e^{-r\mu} \ \ \ \mbox{for all $s,k$}.
$$
Since $x^j(u) = \psi_j(u,0)$, this proves the lemma.

\begin{lemma}
$$
 \frac{\partial x^j}{\partial u^{\alpha}}(u(\lambda x))=
 \lambda \frac{\partial x^j}{\partial u^{\alpha}}(u(x)), \ \ \
 \frac{\partial u^{\alpha}}{\partial x^j}(\lambda x)=
 \frac{1}{\lambda} \frac{\partial u^{\alpha}}{\partial
 x^j}(x).
$$
\end{lemma}

{\it Proof.} It follows from Lemma 1 that
$$
\frac{\frac{\partial x^j}{\partial u^{\alpha}}(u^1+\mu,\dots,u^n+\mu)}
{x^j(u^1+\mu,\dots,u^n+\mu)}=
\frac{\frac{\partial x^j}{\partial u^{\alpha}}(u^1,\dots,u^n)}
{x^j(u^1,\dots,u^n)}.
$$
Therefore we have
$$
\frac{\partial x^j}{\partial u^{\alpha}}(u(\lambda x)) =
\frac{\partial x^j}{\partial
u^{\alpha}}(u^1(x)+\mu,\dots,u^n(x)+\mu)=
$$
$$
= \frac{\partial x^j}{\partial u^{\alpha}}(u(x)) \frac{\lambda
x^j(u(x))}{x^j(u(x))}= \lambda \frac{\partial x^j}{\partial
u^{\alpha}}(u(x)), \ \ \ \lambda=e^{-r\mu},
$$
which proves the first assertion of the lemma. Since
$\frac{\partial u^\alpha}{\partial x^j} \frac{\partial
x^j}{\partial u^\beta} = \delta^\alpha_\beta$, the second
assertion follows from the first one. This proves Lemma 2 and
finishes the proof of Theorem 1.

Given a quasihomogeneous solution to the associativity equations
(\ref{e1}) with a constant invertible matrix
$\left(\eta^{\alpha\beta}\right)$, one may expand it to a
non-semisimple Frobenius manifold as it is explained in \S
\ref{algebraiclemma}.

\section{Examples}
\label{examples}

We present a couple of examples. The first of of them is the simplest solution from an infinite family provided
by Theorem 1 and the second example demonstrates that there are many other solutions with such spectral curves and
which are non given by Theorem 1.

{\sc Example 1.} Let $\Gamma$ is formed by two spheres $\Gamma_1$
and $\Gamma_2$ which intersect at a pair of points (see Fig. 1):
$$
\{a,-a\in\Gamma_1\}\sim\{a,-a\in\Gamma_2\}.
$$

\vskip15mm

\begin{picture}(170,100)(-100,-80)
\qbezier(-10,0)(-10,30)(40,30)
\qbezier(40,30)(90,30)(90,0)
\qbezier(90,0)(90,-30)(40,-30)
\qbezier(40,-30)(-10,-30)(-10,0)
\put(35,33){\shortstack{$\Gamma_1$}}

\qbezier(60,0)(60,30)(110,30)
\qbezier(110,30)(160,30)(160,0)
\qbezier(160,0)(160,-30)(110,-30)
\qbezier(110,-30)(60,-30)(60,0)
\put(105,33){\shortstack{$\Gamma_2$}}

\put(-10,0){\circle*{3}}
\put(90,0){\circle*{3}}
\put(60,0){\circle*{3}}
\put(160,0){\circle*{3}}
\put(75,24){\circle*{3}}
\put(75,-24){\circle*{3}}

\put(-25,0){\shortstack{$P_1$}} \put(165,0){\shortstack{$Q_2$}}
\put(95,0){\shortstack{$Q_1$}} \put(45,0){\shortstack{$P_2$}}
\put(63,20){\shortstack{\small{$a$}}}
\put(54,-26){\shortstack{\small{$-a$}}}
\put(85,18){\shortstack{\small{$$}}}
\put(80,-26){\shortstack{\small{$$}}}
\put(60,-50){\shortstack{Fig. 1}}
\end{picture}

The arithmetic genus of $\Gamma$ equals one: $g_a(\Gamma)=1$.

We consider the case when $n=2$ and $l=1$, i.e. the
Baker--Akhiezer function is normalized at one point $r$. We put $r
\in \Gamma_2$ and $\psi_2(r)=1$.

The function $\psi$ takes the form
$$
\psi_1=e^{u^1z_1}f_0(u^1,u^2),\
\psi_2=e^{u^2z_2}\left(g_0(u^1,u^2)+\frac{g_1(u^1,u^2)}{z_2-c}\right)
$$
and the compatibility conditions read $\psi_1(a)=\psi_2(a),
\psi_1(-a)=\psi_2(-a)$.
This implies
$$
 \psi_1=e^{u^1z_1}\left(
  \frac{2a(c-r)e^{au^1+(a-r)u^2}}{(a+c)(a-r)e^{2au^2}-(a+r)(a-c)e^{2au^1}}
  \right),
$$
$$
 \psi_2=e^{u^2z_2}\left(
 \frac{e^{-ru^2}((a-c)e^{2au^1}+(a+c)e^{2au^2})(c-r)}
 {(a+c)(a-r)e^{2au^2}-(a-c)(a+r)e^{2au^1}}+\right.
$$
$$
 \left. \frac{1}{z_2-c}\frac{(a^2-c^2)(r-c)e^{-ru^2}(e^{2au^1}-e^{2au^2})}
 {(a+c)(r-a)e^{2au^2}+(a-c)(a+r)e^{2au^1}}\right).
$$

The differential $\Omega$ is defined by the differentials
$$
\Omega_1=\frac{\beta}{z_1(z_1^2-a^2)}dz_1,\
\Omega_2=\frac{(z_2^2-c^2)} {z_2(z_2^2-a^2)(z_2^2-r^2)}dz_2.
$$
The regularity condition for $\Omega$ take the form
$$
{\rm res}_{a}\Omega_1={\rm res}_{-a}\Omega_1=\frac{\beta}{2a^2}= -
{\rm res}_{a}\Omega_2 = - {\rm res}_{-a}\Omega_2=
-\frac{(a^2-c^2)}{2a^2(a^2-r^2)},
$$
and implies
\beq
\label{e7}
\beta = \frac{c^2-a^2}{a^2-r^2}.
\eeq
To
achieve the Euclidean metric $\eta_{\alpha\beta}=
\delta_{\alpha\beta}$ we assume that $\varepsilon_1^2 =
\varepsilon_2^2$ which is written as
$$
{\rm res}_{Q_1}\Omega_1=-\frac{\beta}{a^2}= {\rm
res}_{Q_2}\Omega_2=-\frac{c^2}{r^2a^2}
$$
from which we derive that
\beq
\label{e8}
\beta = \frac{c^2}{r^2},\ \ r=\frac{a}{\sqrt{2-\frac{a^2}{c^2}}}.
\eeq
By (\ref{e7}) and (\ref{e8}), we have
the formula which restores $r$ from free parameters $a$ and $c$:
$$
r=\frac{a}{\sqrt{2-\frac{a^2}{c^2}}}.
$$
To obtain real-valued functions $x^1,\dots,x^n$ we have to assume
that $\tau^\ast(\Omega) = \bar{\Omega}$ for $\tau: z_i \to
\bar{z}_i, i=1,2$. This takes place when
$$
a^2,c^2,r^2 \in \R.
$$
The prepotential takes the form
$$
F_{a,c}(x^1,x^2) = \frac{1}{4ac}\left(2x_2\sqrt{(a^2-c^2)x_1^2
+c^2x_2^2}\right.
$$
$$
+2cx_1^2\log \left(-\frac{cx_2+\sqrt{(a^2-c^2)x_1^2+c^2x_2^2}}
{x_1}\right)-\sqrt{2c^2-a^2}(x_1^2+x_2^2)
$$
$$
\left.\times\log\left(c^2(x_1^2-3x_2^2)+a^2(x_2^2-x_1^2)-
 2x_2\sqrt{2c^2-a^2}\sqrt{(a^2-c^2)x_1^2+c^2x_2^2}\right)\right)
$$
and satisfies the associativity equations with
$\eta_{\alpha\beta}=\delta_{\alpha\beta}$.

For $a=1, \ c=\frac{2}{\sqrt{7}}$ the formulas for coordinates and
correlators are rather simple:
$$
x^1=\frac{4(7-\sqrt{7})e^{u^1-u^2}}{(21-6\sqrt{7})e^{2u^1}+(7+2\sqrt{7})e^{2u^2})},
$$
$$
x^2=\frac{e^{-2u^2}(3(\sqrt{7}-3)e^{2u^1}+(5+\sqrt{7})e^{2u^2})}
{3(\sqrt{7}-2)e^{2u^1}+(2+\sqrt{7})e^{2u^2}},
$$
$$
 c_{111}=-\frac{9x_1^8+51x_1^6x_2^2+88x_1^4x_2^4+(2x_1^2x_2^3+4x_2^5)\sqrt{(3x_1^2+4x_2^2)^3}+48x_1^2x_2^6}
 {2x_1(3x_1^4+7x_1^2x_2^2+4x_2^4)^2},
$$
$$
 c_{112}=\frac{9x_1^6x_2+15x_1^4x_2^3-8x_1^2x_2^5+(2x_1^2x_2^2+4x_2^4)\sqrt{(3x_1^2+4x_2^2)^3}-16x_2^7}
 {2(3x_1^4+7x_1^2x_2^2+4x_2^4)^2},
$$
$$
 c_{122}=-\frac{9x_1^7+15x_1^5x_2^2-8x_1^3x_2^4+(2x_1^3x_2+4x_1x_2^3)\sqrt{(3x_1^2+4x_2^2)^3}-16x_1x_2^6}
 {2(3x_1^4+7x_1^2x_2^2+4x_2^4)^2},
$$
$$
 c_{222}=\frac{-27x_1^6x_2-16x_2^7-72x_1^2x_2^5+(4x_1^2x_2^2+2x_1^4)\sqrt{(3x_1^2+4x_2^2)^3}-81x_1^4x_2^3}
 {2(3x_1^4+7x_1^2x_2^2+4x_2^4)^2}.
$$

{\sc Example 2.} Let  $\Gamma$ be the same as in Example 1. In
difference with Example 1 we assume that
$$
P_1 = \infty \in \Gamma_1, \ \ P_2 = 0\in \Gamma_1, \ \ Q_1 =
\infty \in \Gamma_2, \ \ Q_2 = 0 \in \Gamma_2,
$$
the normalization point $R=r$ lies in $\Gamma_1$ and the divisor
of poles $D=c$ lie in $\Gamma_2$ (see Fig. 2). Therewith we do not
assume that the intersection points have the same coordinates:
$$
a \sim b, \ \ -a \sim -b, \ \ \ \pm a \in \Gamma_1, \ \pm b \in
\Gamma_2, \ a \neq b.
$$

\vskip15mm

\begin{picture}(170,100)(-100,-80)
\qbezier(-10,0)(-10,30)(40,30) \qbezier(40,30)(90,30)(90,0)
\qbezier(90,0)(90,-30)(40,-30) \qbezier(40,-30)(-10,-30)(-10,0)

\put(35,33){\shortstack{$\Gamma_1$}}

\qbezier(60,0)(60,30)(110,30) \qbezier(110,30)(160,30)(160,0)
\qbezier(160,0)(160,-30)(110,-30) \qbezier(110,-30)(60,-30)(60,0)

\put(105,33){\shortstack{$\Gamma_2$}}

\put(-10,0){\circle*{3}} \put(90,0){\circle*{3}}
\put(60,0){\circle*{3}} \put(160,0){\circle*{3}}
\put(75,24){\circle*{3}} \put(75,-24){\circle*{3}}

\put(-25,0){\shortstack{$P_1$}} \put(165,0){\shortstack{$Q_2$}}
\put(95,0){\shortstack{$P_2$}} \put(45,0){\shortstack{$Q_1$}}
\put(63,20){\shortstack{\small{$a$}}}
\put(54,-26){\shortstack{\small{$-a$}}}
\put(85,18){\shortstack{\small{$b$}}}
\put(80,-26){\shortstack{\small{$-b$}}}
\put(60,-50){\shortstack{Fig. 2}}

\end{picture}

We take the Baker--Akhiezer function in the form
$$
\psi_1=e^{u^1 2z_1+\frac{u^2}{2z_1}}f(u),\ \ \psi_2 =
g_0(u)+\frac{g_1(u)}{z_2-c}.
$$
The differential $\Omega$ is defined by the differentials
$$
\Omega_1=\frac{z_1}{(z_1^2-a^2)(z_1^2-r^2)}dz_1, \ \ \Omega_2=
\frac{(z_2^2-c^2)} {z_2(z_2^2-b^2)}dz_2.
$$
We have the regularity condition:
$$
\res_{a}\Omega_1= \res_{-a}\Omega_1 = \frac{1}{2(a^2-r^2)}=-
\res_{b}\Omega_2=- \res_{-b}\Omega_2 =-\frac{(b^2-c^2)}{2b^2},
$$
and the Euclidean condition: $\varepsilon^2_1 = \varepsilon^2_2$:
$$
\res_{Q_1} \Omega_2 = -1 = \res_{Q_2}\Omega_2= \frac{c^2}{b^2}.
$$
These conditions are satisfied if and only if $b = \pm ic$ and $a^2
- r^2 = -\frac{1}{2}$. We put
$$
b=i, \ c=-1,\ a =\frac{i}{2},\ r=\frac{1}{2}
$$
and obtain
$$
x^1 = e^{-u^1-u^2}(\cos(u^1-u^2) + \sin(u^1-u^2)), \ \
$$
$$
x^2 = e^{-u^1-u^2}(\cos(u^1-u^2) - \sin(u^1-u^2)).
$$
This gives us the Darboux--Egoroff metric
$$
ds^2 = \left(dx^1\right)^2 + \left(dx^2\right)^2 = 4
e^{-2(u^1+u^2)}\left(\left(du^1\right)^2 + \left(du^2\right)^2
\right)
$$
and a quasihomogeneous solution to the associativity equations
(\ref{e1}) because as in the case of Theorem 1 we have
$$
x^i(u^1+\mu,u^2+\mu) = e^{-2\mu}x^i(u^1,u^2), \ \ \ i=1,2.
$$
Indeed, this solution is very simple and the prepotential $F(x^1,x^2)$
equals
$$
F(x^1,x^2) =
-\frac{1}{8}\left(\left(x^1\right)^2+\left(x^2\right)^2\right)
\log\left(\left(x^1\right)^2+\left(x^2\right)^2\right).
$$
Moreover it is included in a linear pencil of quasihomogeneous functions
$$
F_q(x^1,x^2) = q \left(\left(x^1\right)^2+\left(x^2\right)^2\right)\arctan\left(\frac{x^1}{x^2}\right)
$$
$$
-\frac{1}{8}\left(\left(x^1\right)^2+\left(x^2\right)^2\right)
\log\left(\left(x^1\right)^2+\left(x^2\right)^2\right), \ \ q\in \R,
$$
which satisfy the associativity equations with $\eta_{\alpha\beta}=\delta_{\alpha\beta}$.

The correlators for $F$ are very simple:
$$
c_{111} = -\frac{3}{2}\frac{x^1}{\left(x^1\right)^2+\left(x^2\right)^2} +
\frac{\left(x^1\right)^3}{\left(\left(x^1\right)^2+\left(x^2\right)^2\right)^2}, \ \
$$
$$
c_{112} =
-\frac{1}{2}\frac{x^2}{\left(x^1\right)^2+\left(x^2\right)^2} +
\frac{\left(x^1\right)^2 x^2
}{\left(\left(x^1\right)^2+\left(x^2\right)^2\right)^2},
$$
and the formulas for $c_{122}$ and $c_{222}$ are obtained from the
previous ones by permutation of indices $1 \leftrightarrow 2$.

\section{An algebraic lemma}
\label{algebraiclemma}

\begin{lemma}
Let $F(t^1,\dots,t^n)$ be a solution to the associativity equations with the constant metric $\eta_{\alpha\beta}$.
Then the function
$$
\widetilde{F}(t^0,t^1,\dots,t^n,t^{n+1}) = \frac{1}{2}\left(\eta_{\alpha\beta}t^\alpha t^\beta t^0 +
\left(t^0\right)^2 t^{n+1}\right) + F(t^1,\dots,t^n)
$$
satisfies the associativity equations (\ref{e1}) with the metric
$
\widetilde{\eta} =
\left( \begin{array}{ccc}
0 & 0 & 1 \\
0 & \eta & 0 \\
1 & 0 & 0
\end{array}
\right) $ and the associative algebra generated by
$e_0,e_1,\dots,e_n,e_{n+1}$ with the multiplication law
$$
e_i \cdot e_j = c^k_{ij} e_k, \ \ \ c^k_{ij} = \widetilde{\eta}^{kl}
\frac{\partial^3 \widetilde{F}}{\partial t^l \partial t^i \partial t^j},
$$
has the unity $e_0$:
$$
e_o \cdot e_k =e_k \ \ \ \mbox{for all $k=0,\dots,n+1$},
$$
and the nilpotent element
$e_{n+1}$:
$$
e_{n+1}^2 = 0.
$$
Moreover if $F$ is quasihomogeneous
and $d_\alpha + d_\beta =c$ for all $\alpha,\beta$ such that
$\eta_{\alpha\beta} \neq 0$ then $\widetilde{F}$ is also quasihomogeneous with $d_0 = d_F-c, d_{n+1}=
2c - d_F$ and the same values of $d_\alpha$, $\alpha=1,\dots,n$, as for $F$.
\end{lemma}

The proof of this lemma is straightforward.

Applying this procedure to the examples from \S \ref{examples}, we
obtain four-dimen\-sional Frobenius manifolds $M$ with coordinates
$t^0, t^2=x^1,t^2=x^2, t^3$. The element $e_0$ serves as the unity
and the element $e_3$ is nilpotent in any tangent algebra $T_tM$ :
$e_{n+1}^2 = 0$. In these examples we have $d_F =2, d_1=d_2=1$ and
therefore $d_0=0$ and $d_3=2$.

These examples give two-dimensional deformations of the cohomology
ring of $\C P^2 \sharp \C P^2$. Indeed we have the standard
generators $e_0,\dots,e_3$ in $H^\ast(\C P^2 \sharp \C P^2;\C)$:
$e_0 \in H^0, e_1, e_2 \in H^2, e_3 \in H^4, e_1^2 = e_2^2 = e_3,
e_1e_2=0$. We also have the identity $d_i = \frac{\deg e_i}{2}$.
These deformations change the multiplication rules for
two-dimensional classes by adding two-dimensional terms: $e_i e_j
= e_3 + c_{ij}^k(t) e_k, i,j=1,2$.

We remark that in the Seiberg--Witten theory the associativity
equations also appear even in a more general setting: the matrix
$\eta$ is not necessarily constant and the quasihomogeneity
condition is lifted \cite{MMM}.

\end{document}